\documentclass[%
 reprint,
 amsmath,amssymb,
 aps,
]{revtex4-2}

\usepackage{soul}
\usepackage{graphicx}
\usepackage{dcolumn}
\usepackage{bm}
\usepackage{hyperref}
\hypersetup{
    colorlinks=true,
    linkcolor=blue,
    filecolor=magenta,      
    urlcolor=cyan,
    pdftitle={Logical error estimation from syndrome data of surface-code experiments},
    pdfpagemode=FullScreen,
    }

\begin{document}


\title{Probing the real-time decoding accuracy bounds in trapped ion architectures}

\author{Arianna Meinking}
\email{arianna.meinking@duke.edu}
\affiliation{Duke Quantum Center, Duke University, Durham, NC 27701, USA}
\affiliation{Department of Physics, Duke University, Durham, NC 27708, USA}

\author{Evangelia Takou}
\email{evangelia.takou@duke.edu}
\affiliation{Duke Quantum Center, Duke University, Durham, NC 27701, USA}
\affiliation{Department of Electrical and Computer Engineering, Duke University, Durham, NC 27708, USA}

\author{Kenneth R. Brown}
\email{kenneth.r.brown@duke.edu}
\affiliation{Duke Quantum Center, Duke University, Durham, NC 27701, USA}
\affiliation{Department of Electrical and Computer Engineering, Duke University, Durham, NC 27708, USA}
\affiliation{Department of Physics, Duke University, Durham, NC 27708, USA}
\affiliation{Department of Chemistry, Duke University, Durham, NC 27708, USA}

\date{\today}

\begin{abstract}
Fault tolerant quantum computers will rely on real-time decoders that process syndromes faster than the QEC cycle duration. The millisecond cycle times offered by trapped ion platforms relax low-latency decoding demands, but simulations and experiments have not yet exploited this time budget to reach maximum logical error suppression.
We show how a hybrid sliding window decoding scheme composed of a lower-accuracy and a higher-accuracy decoder can yield substantially improved logical error rates, while preserving low latency. All windows are handled by the lower-accuracy decoder, with some windows being infrequently switched to the higher-accuracy decoder. We extract soft-output information from the lower-accuracy decoder efficiently for any QLDPC code, which allows us to identify decoding windows of low confidence. We simulate memory experiments of bivariate bicycle (BB) codes, using trapped-ion informed noise models. Depending on the physical error rate and BB code, we find up to 3-4 orders of magnitude logical error rate reduction compared to the \texttt{BP-LSD} decoder, and up to 2-3 orders of magnitude lower logical error rate compared to the \texttt{Beamsearch} decoder. By leveraging available classical processing time, our decoding protocol significantly boosts decoding accuracy, ultimately reducing the  demands on increasing code size.


\end{abstract}

\maketitle

\section{Introduction}

Realizing quantum algorithms that surpass classical capabilities requires large-scale fault-tolerant architectures that utilize Quantum Error Correction (QEC). QEC poses hardware constraints which demand that the physical error rates are below a threshold so that logical errors are suppressed exponentially with increasing code size~\cite{aharonov1996arxiv,PreskillJMathPhys2002}. At the same time, QEC relies on the successful inference of errors via a decoding algorithm. For a decoder to be practical, it must be able to decode the syndrome and communicate the correction within the syndrome generation time, to avoid the so-called backlog problem~\cite{TerhalRevModPhys2015}. While for Clifford circuits one can keep track of the Pauli frame, thus avoiding active error-correction, circuits with $T$-gates involve corrective Clifford gates that are conditional on the measurement outcome. 
Deciding which Clifford correction to apply depends on the Pauli frame, meaning that syndromes have to be decoded prior to the classical decision.
Thus, real-time decoding is a necessary requirement for universal logical circuits.

Higher decoder accuracy often comes at the cost of higher decoder latency. For trapped ion and neutral atom quantum computers, 
QEC cycle times range from a few to tens of milliseconds~\cite{PreskillArxiv2026,StutzPRX2021,KimArxiv2025,PoolePRA2025}, whereas for superconducting qubits a QEC cycle lasts around $1~\mu$s~\cite{GoogleNature2025}, posing more stringent requirements for decoders. As such, trapped ion and neutral atom platforms potentially allow for more accurate decoders compared to superconducting platforms, which require very low-latency and FPGA-based decoding methods.   

Quantum low-density parity check (QLDPC) codes, have emerged as more efficient QEC codes, due to their high encoding rates~\cite{BreuckmannPRXQ2021,TremblayPRL2022,PreskillArxiv2026}. Certain QLDPC codes have thresholds comparable to surface codes~\cite{BravyiNature2024}, making them ideal for near-term experimental implementation since they also incur a lower qubit overhead. While these codes are promising, especially for flexible architectures that support certain long-range connections~\cite{Delfossearxiv2026,DelfosseArxiv2026BreakEven,LukinNatPhys2024}, performing logical gates~\cite{YoderArxiv2025} or decoding them~\cite{VasicQuantum2021} is far more complicated than surface codes, and determining accurate decoders that would be viable for real-time implementation is still an open line of research.

Certain decoders are implementable on FPGAs~\cite{HailesIEEE2016,VallsIEEE2021,ZhongIEEE2024,BarberNatElectron2025}, and meet the real-time requirements of QEC. Examples include Belief Propagation (\texttt{BP}) and variants and union-find  (\texttt{UF})~\cite{DelfosseQuantum2021}. BP alone does not produce a threshold for most QEC codes, so it is often modified or combined with other decoders to improve performance. Specifically, the clustering decoder \texttt{BP-LSD}~\cite{RoffeNatCommun2025} acts as a lower-latency alternative to \texttt{BP-OSD}~\cite{PanteleevQuantum2021}, while retaining the logical error rate improvements of \texttt{OSD}. Some variants of \texttt{BP} are equipped to handle erasures; namely, the \texttt{BPGD} is a promising \texttt{BP} variant for state-of-the-art erasure-aware decoding ~\cite{gokduman_erasure_2024}. \texttt{UF} handles erasure and Pauli errors most effectively in topological codes~\cite{DelfosseQuantum2021}, and has also been adapted for QLDPC codes~\cite{Sorensenarxiv2024,DelfosseIEEE2022}, yielding lower thresholds compared to topological codes. While the aforementioned decoders achieve low-latency, they cannot reach very high decoding accuracy, especially for QLDPC codes whose decoding structures are not matchable.

Recently, \texttt{Relay-BP} ~\cite{Mullerarxiv2025} was proposed as a potentially fast and FPGA-friendly decoder, which deals with trapping sets encountered in the decoding of QLDPC codes with BP. \texttt{Relay-BP}  was implemented on FPGA and was benchmarked on synthetic data~\cite{maurerArxiv2025}. A recent implementation of \texttt{Relay-BP}  on a GPU~\cite{HongArxiv2026} resulted in average latencies of a few milliseconds, and maximum latencies of hundreds of milliseconds, which are more appropriate for trapped ion architectures than superconducting qubits. Further, it was also found that \texttt{Relay-BP}  can have suboptimal performance at low noise due to trapping sets. To reduce the decoding latency of \texttt{Relay-BP}, a fully-parallelized FPGA implementation would be needed, whereas further exploration of its accuracy in the presence of realistic hardware noise is yet to be demonstrated. Additionally, \texttt{Relay-BP}  requires long-term memory which could lead to long runtime tail distributions and could pose challenges for FPGA implementations.

In this work, we propose a real-time decoding protocol for QLDPC codes tailored to noise present in trapped ion architectures. We combine sliding window decoding with decoder switching: an idea developed in Ref.~\cite{Fujiiarxiv2025} and tested on surface codes using matching-based decoders. By switching from a weak (low-accuracy) and low-latency decoder to a higher-accuracy, but slower decoder as needed, the decoding accuracy can be boosted, while avoiding the syndrome backlog. We determine switch rates that are feasible for trapped ion architectures, by employing the cluster norm soft output information~\cite{BartlettnpjQI2026}. This metric can be calculated efficiently for QLDPC codes, avoiding the exponential cost incurred by other metrics such as logical or complementary gaps which require decoding of each logical class. Our protocol switches between \texttt{BP-LSD} or \texttt{UF} weak decoders to the \texttt{Tesseract} or \texttt{Relay-BP} strong decoders. We exploit the longer QEC cycle times offered by trapped ion architectures to improve the logical error rates of memory experiments on bivariate bicycle (BB) codes. Our main results are summarized as follows:
\begin{itemize}
    \item Under Pauli errors, decoder switching yields 
    a logical error rate improvement that  
    even exceeds
    3 orders of magnitude  compared to \texttt{BP-LSD}, for  $p=2\times 10^{-3}$.
    \item Comparing against the \texttt{Beamsearch} decoder~\cite{YePRXQuantum2026} under Pauli noise, our protocol achieves up to 2-3 orders of magnitude lower logical error rates for $p=2\times 10^{-3}$.
    \item For Pauli+leakage errors, we find that \texttt{Relay-BP} becomes an unreliable strong decoder, and compromises the success of the decoder switching protocol, if no leakage reduction methods are employed. On the contrary, when the \texttt{Tesseract} decoder is selected as the strong decoder, the performance gains of the decoder switching protocol persist in the presence of leakage.
    \item For Pauli+erasure errors, we find that decoder switching yields up to 3 orders of magnitude logical error suppression compared to Pauli-only errors, when the \texttt{UF} is chosen as the weak decoder. Since \texttt{UF} is a very fast decoder, decoder switching preserves a very favorable decoding latency.
    \item Importantly, the decoding latency of our protocol is comparable or better than the \texttt{Beamsearch} decoder, and the performance gains are estimated based on experimentally achievable switch rates for trapped ion architectures. With decreasing physical error rates, the requirements of decoder switching drop, as one can switch less frequently to achieve appreciable logical error rate gains.
\end{itemize}




While we assess our decoder run-times in terms of the theoretical bound imposed by the switch rate between weak and strong decoding, we also compare real run-times of weak and strong decoders on a CPU. We find that although \texttt{BP-LSD} is expected to have almost linear runtime ($\mathcal{O}(n\alpha(n))$) in the subthreshold regime ~\cite{RoffeNatCommun2025}, when tested on a local CPU, it is in practice faster than \texttt{Tesseract} for $p\geq 5\times 10^{-4}$. This is because the proposed fully parallel and optimized algorithm of Ref. ~\cite{RoffeNatCommun2025} has not been implemented. Because the proposed algorithm uses a parallelized version of the union-find data structure, we expect the runtime to be similar to the cost of \texttt{UF}. Since our decoder switching protocol that uses \texttt{UF} as a weak decoder is feasible in real-time, we expect that improving the implementation of \texttt{BP-LSD} will result in similar latencies, but this remains to be shown. 


The package we developed and all our codes that reproduce the results of this paper are provided in~\cite{realtimeGithub}.

This paper is organized as follows. 
In Sec.~\ref{subsec:decoding_overview} we discuss prior work and introduce the decoding protocol for QLDPC codes. In Sec.~\ref{sec:decoder_characterization} we characterize the logical performance of decoders we select and determine the switch rates for each BB code. In Sec.~\ref{Sec:Results} we benchmark the performance of the decoder switching protocol under Pauli, Pauli+leakage and Pauli+erasures noise models. Finally, in Sec.~\ref{Sec:Discussion} we conclude our work and discuss future directions.

\section{Overview of decoding protocol \label{subsec:decoding_overview}} 

Our goal is to develop a viable real-time decoding protocol, tailored to trapped ion QEC architectures. However, the ideas presented below are also applicable to other quantum computing platforms. For trapped ions, as well as neutral atoms, typical QEC cycle times range from a few to tens of milliseconds~\cite{StutzPRX2021,KimArxiv2025,PreskillArxiv2026,PoolePRA2025,ChongIEEE2025}. We use this to our advantage to boost the decoding accuracy of memory experiments of QLDPC codes, while preserving a latency that aligns with the demands of trapped ion platforms. 

{\color{black}{In what follows, we discuss previous work, and then introduce the decoding procedure we implement.

\subsection{Prior work \label{subsec:prior_work}}

Real-time decoding requires that syndromes are processed as soon as possible after their generation, without waiting for the complete syndrome history to become available. This is addressed by sliding window decoding~\cite{PreskillJMathPhys2002, CampbellNatCommun2023,KangQuantum2025}, which processes the spacetime decoding volume in smaller regions, typically sliced across the time dimension. Rather than decoding the full syndrome history, syndromes are decoded across $W$ QEC cycles. From these syndromes, only those that span $F$ cycles are committed, which belong in the so-called commit region. The window of thickness $W$ then slides forward in time such that the overlap between the current and past window has $F$ cycles. The $W-F$ region is known as the buffer region, and has size $n_{\text{buffer}}$. The size of the buffer region is important for preserving code distance;  a buffer window size of $\mathcal{O}(d)$ cycles~\cite{Bombiarxiv2023,CampbellNatCommun2023} is necessary, so that error chains of weight $\leq d/2$ are not mistaken for logical errors. In some cases, the window size, and hence buffer size, is kept smaller than $\mathcal{O}(d)$, at the expense of a higher logical error rate.

While sliding window decoding bounds the spacetime volume of the decoding task, there is still a trade-off between decoding accuracy and speed, since the decoding time needs to be strictly smaller than the syndrome generation time. Recently, Ref.~\cite{Fujiiarxiv2025} proposed a decoder switching protocol, where each  window is first handled by a weak (less accurate) decoder of low latency. When the weak decoder is not confident on a given syndrome, the window is switched to a strong (more accurate) decoder. Soft-output information from the weak decoder reveals the decoder confidence, and an appropriate cutoff determines the rate at which windows are switched to the strong decoder. Since switching events to the strong (and slower) decoder do not happen very often, the requirement for avoiding the syndrome backlog is relaxed based on the switch rate, allowing for increased decoding accuracy. This work applied the decoding switching protocol on surface codes undergoing Pauli noise, and focused only on matching-based decoders, namely the \texttt{UF}~\cite{DelfosseQuantum2021}, \texttt{MWPM}~\cite{HiggottACM2022} and  \texttt{Belief-Matching}~\cite{HiggottPRX2023} decoders. As surface codes encode a single logical qubit, the decoder confidence was  efficiently obtained from the weak decoders in Ref.~\cite{Fujiiarxiv2025} via the complementary gap~\cite{ Gidney2024YokedSurfaceCodes}. However, for QLDPC codes, calculating logical~\cite{BombinPRXQ2024} or complementary gaps scales exponentially with the number of logical qubits as it requires decoding over an exponential number of logical classes. Therefore, an efficient method to extract soft-output information from a decoder is needed for QLDPC codes. Further, the viability of this protocol for QLDPC codes, for which decoding is less trivial, was not explored.

\subsection{Extension to QLDPC codes \label{subsec:qldpc_switching}}

In our work, we design the decoder switching protocol with QLDPC codes in mind. To efficiently calculate decoder confidence from the weak decoders, we employ the cluster norm metric, which was developed and used for post-selection purposes in Ref.~\cite{BartlettnpjQI2026}. Here we utilize this metric to determine which windows need to be handled by the strong decoder. The cluster norm can be obtained from clustering-based decoders, such as \texttt{UF}, \texttt{BP-LSD}, and ambiguity clustering~\cite{WolanskiIEEE2024}, and in this work we will consider the \texttt{UF} and \texttt{BP-LSD} decoders.

The intuition behind the cluster norm is that larger error clusters in the decoding structure indicate higher probability of logical errors. The error clusters correspond to corrections found during the course of decoding. If the clusters are large, then it is more likely they overlap with logical operators. Thus, the ratio of cluster norm to the total number of error mechanisms in a decoding region gives a reasonable heuristic metric of confidence. The cluster norm is given by~\cite{BartlettnpjQI2026}:

\begin{equation}
    Q^{(\alpha)} = \frac{1}{|\mathcal{E}|}\left( \sum_i |C_i|^\alpha\right)^{1/\alpha},
    \label{eq:cluster_norm}
\end{equation}
where $C_i$ is the $i$-th cluster in the set of clusters found during decoding, and $\mathcal{E}$ is the full set of spacetime error mechanisms (faults). 
For the purposes of this work, we set $\alpha = 2$. Since we decode with sliding window, it is necessary to restrict the clusters only within corrections we accept, namely within the commit region. As a larger cluster norm indicates a higher probability of logical error, we impose switching cutoffs such that the weak decoder's correction is rejected if $Q^{(2)} >q_{th}$. If the cluster norm exceeds this cutoff in a given window, we discard the weak correction and switch to strong decoding. 

The cutoffs we impose depend on the noise model, as well as characteristics of the platform, since the switch rates that will eventually be practical depend on the syndrome generation time. We will revisit this, when we specifically discuss platform suitability, and estimate the logical error rate gains for trapped ion platforms.

\subsection{Syndrome extraction circuits and noise models \label{Sec:NoiseModel}}
In all following simulations, we construct our memory syndrome extraction circuits  using the \texttt{QUITS} package \cite{KangQuantum2025}, and we sample the circuits using Stim~\cite{GidneyQuantum2021}. Our framework is generically applicable to QLDPC codes, but we focus on the  [[72, 12, 6]], [[90, 8, 10]], [[126, 8, 10]], [[144, 12, 12]], and [[162, 8, 14]] BB codes, whose parameters are given in~\cite{BravyiNature2024,MuellerQuantum2026}. 

Our trapped ion noise models are based on the recent architecture blueprint of IonQ~\cite{Delfossearxiv2026}. The first model we explore is the fully connected trapped-ion noise model of Ref.~\cite{Delfossearxiv2026}, which is a purely depolarizing noise model with no additional transport step, as described in Table~\ref{tab:IonicPauli}. 

\begin{table}[!htbp]
    \centering
    \begin{tabular}{|c|c|}
    \hline
         Preparation & $p/10$  \\
         \hline
         Single-qubit gate & $p/10$ \\
         \hline
         Single-qubit gate & $p/10$ \\
         \hline
         Two-qubit gate & $p$ \\
         \hline
         Measurement & $p/10$ \\ 
         \hline
         Idle & $p/100$  \\
         \hline
    \end{tabular}
    \caption{Fully connected trapped-ion noise model. The first column corresponds to a circuit component, and the second column to the depolarizing rate.}
    \label{tab:IonicPauli}
\end{table}

Although depolarizing noise is widely adopted in the literature due to its simplicity, leakage, loss or erasure errors are equally important. Despite their typically smaller magnitude compared to depolarizing noise, they can introduce correlated errors as they propagate through two-qubit gates. To determine the viability of the decoder switching protocol, we will thus consider experimentally informed leakage and erasure rates.

}}

{\color{black}{
\subsection{Choice of decoders}
\label{subseq:choice_of_decoders}
The \texttt{UF} and \texttt{BP-LSD} weak decoders we choose are ideal for two reasons: i) they allow for efficient calculation of cluster norms, and ii) they satisfy the low-latency requirements for real-time decoding. Additionally, the \texttt{UF} algorithm is ideal for handling erasures, and in both trapped-ion and neutral atom architectures leakage from computational states can be converted with high efficiency to erasures~\cite{KangPRXQ2023,ThompsonNatPhys2026}.

We frame the decoding problem using a Detector Error Model (DEM) and spacetime parity check matrix $\mathbf{H}$ mapping the independent physical fault mechanisms to detectors in the DEM. Given a parity check matrix $\mathbf{H}$ and an observed detector error syndrome $\mathbf{s}$, both \texttt{BP-LSD} and \texttt{UF} seek to solve the decoding problem $\mathbf{H}\cdot \mathbf{e} = \mathbf{s}$ by growing candidate error clusters from the known syndrome until validity \cite{DelfosseIEEE2022,RoffeNatCommun2025}. 

The \texttt{BP-LSD} algorithm begins by completing a pre-selected number of Belief Propagation (\texttt{BP}) iterations using the error syndrome. The resulting correction and DEM weights are stored, and become the soft information input for the LSD part. LSD grows error clusters  incrementally by adding the most likely error mechanisms (weighted expansion) to a local parity check sub-matrix \cite{RoffeNatCommun2025}, to make the clusters valid and satisfy the local decoding problem. Because in the sub-threshold regime errors belong in disjoint decoding regions, \texttt{BP-LSD} exploits the sparsity to solve the inversion problem $e=H^{-1} s$ for each localized cluster in parallel.

\begin{figure*}[!htbp]
    \centering
    \includegraphics[scale=0.76]{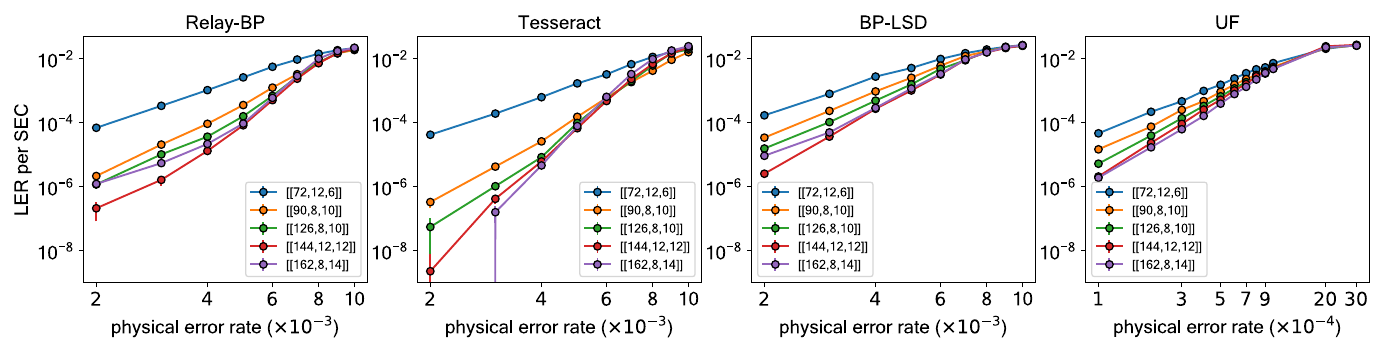}
    \caption{
    Decoding performance for strong and weak decoders, under sliding window decoding. Each panel shows the logical error rate (LER) per syndrome extraction cycle (SEC), for five BB codes and different decoders, as a function of the physical error rate. The commit region for each code is set to $F=d/2$, where $d$ is the distance of the code, the buffer region is set to $n_{\text{buffer}}=d$, with a total window size of $W=F+n_{\text{buffer}}$. From left to right, we display the performance of \texttt{Relay-BP}, \texttt{Tesseract}, \texttt{BP-LSD} and \texttt{UF} decoders.}
 \label{fig:Sliding_window_benchmarks}
\end{figure*}

Similarly, \texttt{UF} grows clusters from an error syndrome. However, rather than prioritizing the addition of single most likely faults, unweighted \texttt{UF} grows error clusters by uniformly adding all adjacent error mechanisms to a check node \cite{Sorensenarxiv2024}. For topological codes, \texttt{UF} validates clusters by ensuring that the parity within the cluster is even, meaning defects can be connected via error chains to give rise to even parity. However, satisfying the parity constraint is not sufficient for general QLDPC codes. For this reason, we use the \texttt{UF} algorithm of Ref.~\cite{Sorensenarxiv2024}, which is adapted to QLDPC codes, and ensures validity of the local decoding problem by solving $\textbf{s} = \textbf{H} \cdot \textbf{e}$ via Gaussian elimination. This \texttt{UF} variant follows an unweighted growth of clusters.

As strong decoders, we select \texttt{Relay-BP} ~\cite{Mullerarxiv2025} and \texttt{Tesseract} \cite{Shuttyarxiv2025, Shuttyarxiv2026} due to their superior logical error rate suppression on BB codes. Also, their decoding latencies on CPU lie in the milliseconds regime [see Appendix~\ref{App:Decoding_times}], which approach the QEC cycle times of trapped ion platforms, and can thus be readily used as CPU-based strong decoders.

The \texttt{Relay-BP}  decoder is a modification of \texttt{BP}. The decoder ensembles $R$ relay `legs', each decoded via \texttt{DMem-BP} \cite{Mullerarxiv2025}. For each leg, $T_r$ \texttt{BP} cycles with additional memory updates indicated by $\Gamma_r$ update the DEM weights. At the end of decoding, $S$ solutions are chosen from the ensemble. \texttt{Relay-BP}  parameters are highly code specific and determine decoding performance. The parameters we choose are those optimized in \cite{Mullerarxiv2025} for the Gross code.

The \texttt{Tesseract} decoder serves as our second strong decoder \cite{Shuttyarxiv2025,Shuttyarxiv2026}. Operating as a Maximum Likelihood Estimation (MLE) decoder, it performs an $A^*$ search on a pruned graph of candidate error configurations to return the globally most likely physical error pattern.

More information about the decoding parameters we set, and the decoding latency is provided in Appendix~\ref{App:DecoderParams} and Appendix~\ref{App:Decoding_times}, respectively.

}}

\section{Decoder characterization under sliding window decoding \label{sec:decoder_characterization}}

\subsection{Logical error rate performance \label{subseq:ler_sliding_window}}

We begin by characterizing the logical error rate (LER) performance of the two strong decoders (\texttt{Relay-BP}, \texttt{Tesseract}) and the two weak decoders (\texttt{BP-LSD}, \texttt{UF}), considered in this work, when the BB codes are decoded with sliding window decoding. We consider $Z$-memory experiments, where the circuits are built using built-in functions from QUITS~\cite{KangQuantum2025}.

To preserve similar logical performance as for $\mathcal{O}(d)$-cycles decoding, we set $n_{\text{buffer}}=d$, and a commit region size of $F=d/2$, giving rise to a total window size of $W=F+n_{\text{buffer}}$, with $d$ being the code capacity distance of each code. The circuit distance, $d_{\text{eff}}$, can be smaller than $d$, but this is not a problem, since we still satisfy $n_{\text{buffer}}\sim\mathcal{O}(d)$, where $d\geq d_{\text{eff}}$. We simulate  $r=25$ syndrome extraction cycles for each code, and calculate the LER per syndrome extraction cycle (SEC), $\epsilon=1-(1-p_L(r))^{1/r}$. Here, $p_L(r)$ is the LER after $r$ syndrome extraction cycles, averaged over the number of logical qubits for each code. We verified numerically that the choice of syndrome extraction cycles does not change the value of $\epsilon$ we extract, and it stays approximately constant. Our choice of $r$ is made such that we can slide at least a few windows for each BB code we study.

In Fig.~\ref{fig:Sliding_window_benchmarks} we show the LER per SEC for each decoder as a function of the physical error rate. The physical error rate corresponds to the two-qubit error rate, based on which we scale the remaining rates in the circuit [see Sec.~\ref{Sec:NoiseModel}]. \texttt{BP-LSD} and \texttt{UF} have larger logical error rates compared to \texttt{Relay-BP} and \texttt{Tesseract}, which is why they are considered weak (but faster) decoders in our work. While \texttt{UF} is the least accurate, it is the fastest decoder of all, and it can handle erasures, a noise source that we will address in a next section. Additionally, the expansion of clusters happens in an unweighted and uniform manner, but further performance gains could be obtained under a weighted \texttt{UF} version [c.f. Ref.~\cite{BrownPRA2020}]. The most accurate decoder, as expected, is \texttt{Tesseract}, since it is an approximate MLE decoder. In our simulations we noticed that \texttt{Relay-BP} is not always converging, especially for larger physical error rates, since it is a heuristic decoder. Thus, in some cases, the window correction obtained by \texttt{Relay-BP} does not satisfy the syndrome, which can corrupt the syndromes in the next window,  potentially leading to an overall logical failure. This is a feature that does not play a very important role when \texttt{Relay-BP} is benchmarked on its own, but needs special handling when decoder switching is considered. This is because the strong decoder could corrupt the syndrome update for the next window handled by the weak decoder. We will address this issue when we consider the decoder switching protocol later on.

\begin{figure*}
    \centering
    \includegraphics[width=\textwidth]{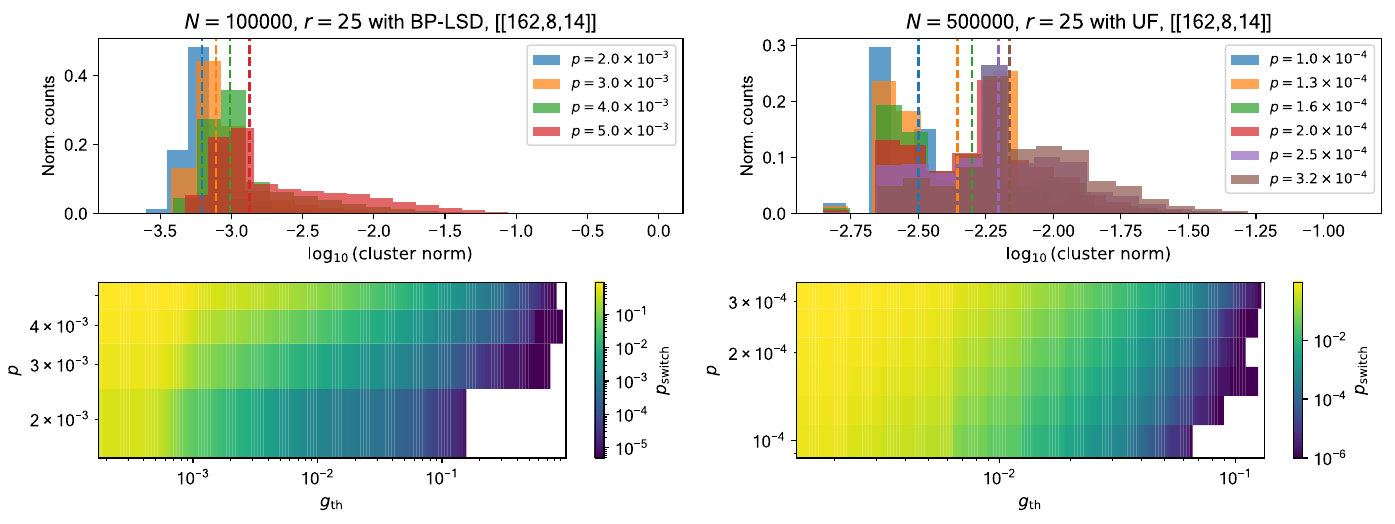}
    \caption{Cluster norm and switch rate characterization for the [[162,8,14]] code, under the Pauli noise model of Table~\ref{tab:IonicPauli}. 
    The top panels show the cluster norm histograms obtained from \texttt{BP-LSD} (left) or \texttt{UF} (right), for varying physical error rates, $p$. The bottom plots show the switch probability by imposing each cutoff at each physical error rate.}
    \label{fig:heatmaps_162_8_14}
\end{figure*}

\subsection{Cluster norm distributions and switch rates \label{Sec:ClusterNorms}}
To determine when to switch a window from weak to strong decoding, we use the cluster norm metric described in Sec.~\ref{subsec:qldpc_switching}. We pre-compute cluster norm distributions for each code and physical error rate using Eq.~\eqref{eq:cluster_norm}. 
By calculating how many cluster norms are above a cutoff (i.e. how often $Q^{(2)} > q_{th}$) over many shots and windows,  we create a cumulative distribution for each code and error probability [see Appendix~\ref{app:cluster_characterization}]. This cluster norm characterization gives us directly the  switch probability $p_s$. We can thus impose informed cutoffs to select $p_s$ based on platform constraints. An example cluster norm distribution for many physical error rates, $p$, and a resultant heatmap from scanning $q_{th}$ are shown in Fig. \ref{fig:heatmaps_162_8_14}. As expected, when we increase the cutoff, $q_{th}$, we accept larger cluster norms (hence, lower decoding confidence), which means that the switch events are rare. Similarly, given a fixed cutoff, as the physical error rate increases we are forced to switch more often to preserve the same decoding confidence.


We note that the cluster norm distributions obtained from \texttt{UF} and \texttt{BP-LSD}  are different. Since \texttt{UF} has lower accuracy than \texttt{BP-LSD}, we consider cluster distributions at lower error rates. Further, unweighted \texttt{UF} (\texttt{BP-LSD}) effectively completes a breadth-first (depth-first) search of the tanner graph to form clusters, which means the \texttt{UF} clusters are larger. These factors impact the switch rates and the performance of the switching protocol. Using a weighted \texttt{UF} could potentially reduce the cluster norms.


\section{Decoder switching performance \label{Sec:Results}}
\subsection{Pauli noise}
\begin{figure*}[!htbp]
    \centering
    \includegraphics[scale=0.8]{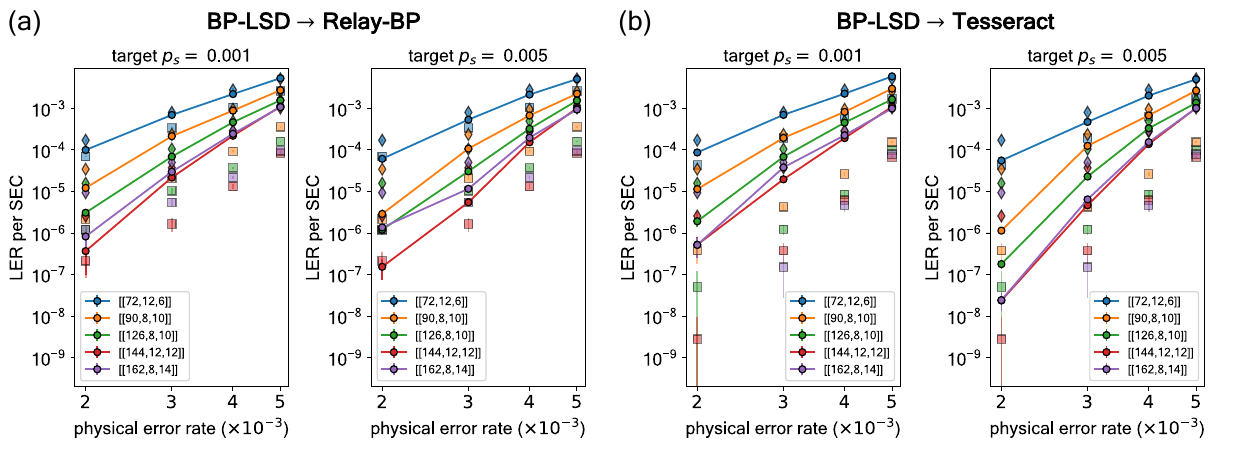}
    \caption{Logical error rate performance of decoder switching, when \texttt{BP-LSD} is chosen as the weak decoder. (a) Logical error rate per syndrome extraction cycle for a target switch rate of $p_s=10^{-3}$ (left), and $p_s=5\times 10^{-3} $ (right), as a function of physical error rate, when the strong decoder is \texttt{Relay-BP} . (b)  Same as in (a) when \texttt{Tesseract} is chosen as the strong decoder. In all panels, the solid lines show the decoder switching performance, squares show the performance of the strong decoder, and diamonds the performance of \texttt{BP-LSD}.}
    \label{fig:DS_bplsd}
\end{figure*}

\begin{figure*}
    \centering
    \includegraphics[width=\textwidth]{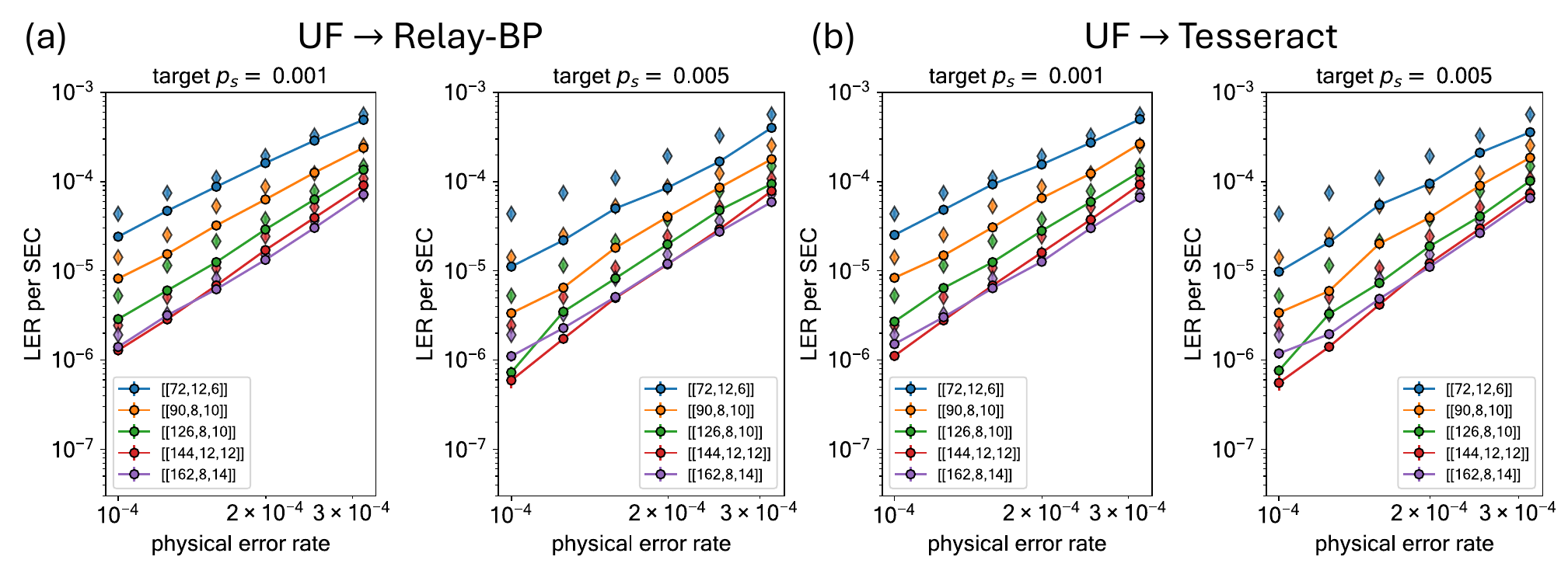}
    \caption{Logical error rate per syndrome extraction cycle of decoder switching, when \texttt{UF} is chosen as the weak decoder. (a) Switching from \texttt{UF} to \texttt{Tesseract}, and (b) switching from \texttt{UF} to \texttt{Relay-BP}. The (left), (right) plots indicate the target switch rates of $ 10^{-3}$ and $5 \times 10^{-3}$, respectively. The diamond markers show weak decoding performance. 
    We note that the physical error rate is lower than the analysis in Fig. \ref{fig:DS_bplsd}, since \texttt{UF} has a lower accuracy than \texttt{BP-LSD}. 
    }
    \label{fig:uf_switching}
\end{figure*}

To benchmark the performance of the decoder switching protocol, we first consider the trapped-ion Pauli noise model of Ref.~\cite{Delfossearxiv2026} [see Sec.~\ref{Sec:NoiseModel}], and explore four combinations of decoder switching:
\begin{itemize}
    \item \texttt{BP-LSD} to \texttt{Relay-BP},
    \item \texttt{BP-LSD} to \texttt{Tesseract}, 
    \item \texttt{UF} to \texttt{Relay-BP},
    \item \texttt{UF} to \texttt{Tesseract}.
\end{itemize}
We fix the switch rate across all codes and physical error rates
to either $p_s=10^{-3}$ or $p_s=5\times 10^{-3}$. The relevant cluster norm cutoffs to preserve our target switch rates are estimated based on the analysis of Sec.~\ref{Sec:ClusterNorms}. Note that because we estimated those cutoffs using 100,000 or 500,000 shots per code and physical error rate, we achieve approximately the target switch rates. For \texttt{BP-LSD} switching, we explore two-qubit rates $p\in[2\times 10^{-3},5\times 10^{-3}]$, whereas for \texttt{UF} switching we probe $p\in [10^{-4},3\times 10^{-4}]$. Specifically, $p=10^{-4}$ has been achieved only for very small trapped ion systems~\cite{HartyPRXQuantum2025,HartyArxiv2025}. For larger systems, such as the 98 trapped-ion quantum processor Helios of the Quantinuum team, the measured average two-qubit error rate was $8\times 10^{-4}$~\cite{QuantinuumNature2026}.

Figure~\ref{fig:DS_bplsd}(a) shows the LER per SEC for \texttt{BP-LSD} to \texttt{Relay-BP} switching and Fig.~\ref{fig:DS_bplsd}(b), for \texttt{BP-LSD} to \texttt{Tesseract} switching. As expected, increasing the switch rate improves the performance of the decoder switching protocol. For $p_s=10^{-3}$, the largest improvement is observed at $p=2\times 10^{-3}$. 
We find that using \texttt{Tesseract} as the strong decoder of the switching protocol suppresses the LER more than \texttt{Relay-BP}, as expected based on their decoding accuracy. This happens for $p_s=5\times 10^{-3}$, but not for $p_s=10^{-3}$, meaning there is a transition point at which invoking the strong decoder more often allows us to obtain extra performance gains.
As we mentioned in Sec.~\ref{subseq:ler_sliding_window}, 
\texttt{Relay-BP} might not converge, as it is a heuristic decoder. Here we handle those failures by checking if the correction satisfies the syndrome, and otherwise we simply accept the solution of the weak decoder. In an actual implementation, provided buffer regions are created both prior and after the
strong decoding window, the solution of the strong decoder could be accepted, and simply merged in the end
with the solutions of the weak decoding regions. An alternative solution would be to discard the strong decoding
correction. 

We also observe performance improvements when using \texttt{UF} as our weak decoder. In Fig. \ref{fig:uf_switching}(a) we show the LER per SEC for \texttt{UF} to \texttt{Relay-BP} switching, and in Fig.~\ref{fig:uf_switching}(b) for \texttt{UF} to \texttt{Tesseract} switching. For both combinations, switching suppresses the LER, and the improvements are more pronounced as the switch rate increases. In particular, switching improves the LER by an order of magnitude compared to \texttt{UF} when $p_s = 5\times 10^{-3}$, and $p =  10^{-4}$. Since the LER  is bounded by the strong and weak decoder performance, switching from \texttt{UF} still has higher LER per SEC than switching from \texttt{BP-LSD}, given the same switch rate. Thus, in this case, $p_s=5\times 10^{-3}$  is not enough to reach the strong decoder performance as we found for \texttt{BP-LSD}. This is because there is a larger mismatch of decoding accuracy between the weak and strong decoders, under Pauli noise model.


Another interesting feature we discover is that the switching performance saturates to the same LER independent of our strong decoder choice, which is in contrast to our conclusions from Fig.~\ref{fig:DS_bplsd}, where switching to \texttt{Tesseract} yielded the lowest LER. 
We believe there are two possible explanations for this feature. One reason could be that the cluster norm metric is not as reliable as for \texttt{BP-LSD}, since the growth of clusters happens in an unweighted way for \texttt{UF}. Thus, the windows that are chosen to be switched (e.g., windows that could cause logical errors) might not align well with decoding uncertainty. Another possibility is that this is again owed to the large accuracy mismatch between \texttt{UF} and strong decoders, and given that per shot we switch at most a few windows to the strong decoders, the shots they can save from logical failures are limited and bounded by \texttt{UF}'s accuracy. Since in the limit of $p_s=1$ all windows are handled by the strong decoder, we believe there exists a $p_s<1$ where switching to \texttt{Tesseract} yields a lower LER than switching to \texttt{Relay-BP}. This phenomenon highlights that the choice of weak and strong decoders is crucial for obtaining maximal performance gains and should not be arbitrary.


\subsection{Platform-based logical error rate improvement}
\label{sec:platform_ler_improvement}

To implement the decoder switching protocol on hardware, we need to consider specific details of the physical platform. Based on 
syndrome collection and decoding times, we can upper bound the switch rates and estimate performance gains that can be realistically achieved.

The decoder switching protocol achieves an average latency per window of 
\begin{equation}
    T_{\text{av.}} = T_w(1-p_s)+p_s (T_w+T_s)=T_w +p_s T_s,
\end{equation}
where $p_s$ is the switch rate and $T_s$ ($T_w$) is the latency of the strong (weak) decoder. This latency includes both decoding time per window, and communication time. To avoid the backlog, the decoding latency has to be less than the syndrome generation time, $t_{\text{gen.}}$, meaning~\cite{Fujiiarxiv2025}:
\begin{equation}
    T_w < t_{\text{gen.}} ~~\text{and}~~p_sT_s<t_{\text{gen.}}.
\end{equation}
If the window consists of $r$ syndrome extraction cycles, then $t_{\text{gen.}}=r\times \tau_{\text{cycle}}$. Assuming the decoding time for a window scales linearly with the number of rounds, we get a weak decoder latency of:
\begin{equation}\label{Eq:DecLatency}
    T_w = r \times (\tau_{w}^{\text{dec.}}+\tau_{w}^{\text{comm.}}).
\end{equation}
A similar definition holds for the strong decoder. Using Eq.~(\ref{Eq:DecLatency}), the conditions to avoid the syndrome backlog take the form:
\begin{equation}
\alpha_{w}:=\frac{(\tau_w^{\text{dec.}}+\tau_{w}^{\text{comm.}})}{\tau_{\text{cycle}}}<1 
\end{equation}
and:
\begin{equation}
    \beta_s:=\frac{p_s (\tau_s^{\text{dec.}}+\tau_s^{\text{comm.}})}{\tau_{\text{cycle}}}<1.
\end{equation}
The advantage of the decoder switching protocol is that the backlog condition for the strong decoder is relaxed by a factor of $p_s$, allowing us to improve decoding accuracy beyond the weak decoder limits.

The choice of switch rates and decoders that satisfy the above conditions depend on $\tau_{\text{cycle}}$, which is specific to the QEC platform. For trapped ions and neutral atoms, typical cycle times are to the order of 1-10~ms, whereas for superconducting qubits cycle times are close to 1~$\mu$s. Thus, fixing the same strong decoder, the switch rate can be $10^3-10^4$ times larger for trapped ions compared to superconducting qubits.

Considering a strong decoding time of $\tau_{s}^{\text{dec.}}=1$~ms, a communication time of $\tau_{s}^{\text{comm.}}=10\tau_{\text{cycle}}$
, and a typical trapped-ion cycle time of $\tau_{\text{cycle}}=1$~ms, implies that the switch rate should satisfy $p_s\lesssim 9.09\times 10^{-2}$. The long communication time is taken here as a worst-case estimate, but depending on specific implementations it could be faster than the decoding time. For superconducting qubits, using a cycle time of $\tau_{\text{cycle}}=1~\mu$s (with $\tau_s^{\text{comm.}}=10\tau_{\text{cycle}}$), requires that $p_s \lesssim 9.9 \times 10^{-4}$. While this strong decoding time does not allow enough room for LER improvement for superconducting platforms, a strong decoding time of 100~$\mu$s yields $p_s \lesssim 10^{-2}$, and could still enable significant LER improvements for these platforms.

\begin{figure}[!htbp]
    \centering
    \includegraphics[scale=1.1]{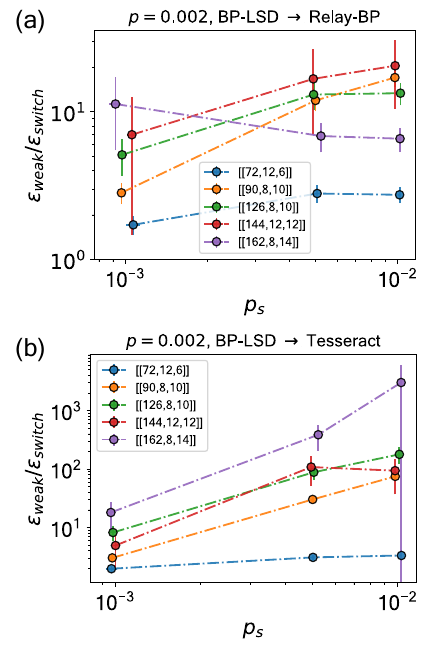}
    \caption{Per-cycle logical error rate ratio of the switching protocol compared to \texttt{BP-LSD}, considering only Pauli-type noise.
    (a) LER ratio of weak to decoder switching protocol as a function of switch rate, when the strong decoder is selected to be \texttt{Relay-BP} . (b) Same as in (a), but selecting  \texttt{Tesseract} as the strong decoder. In both cases, we consider the physical error rate of $p=2\times 10^{-3}$. Error bars correspond to standard deviation and are included  along both axes.}
    \label{fig:Improvement_bplsd}
\end{figure}
\begin{figure}
    \centering
    \includegraphics[width=\linewidth]{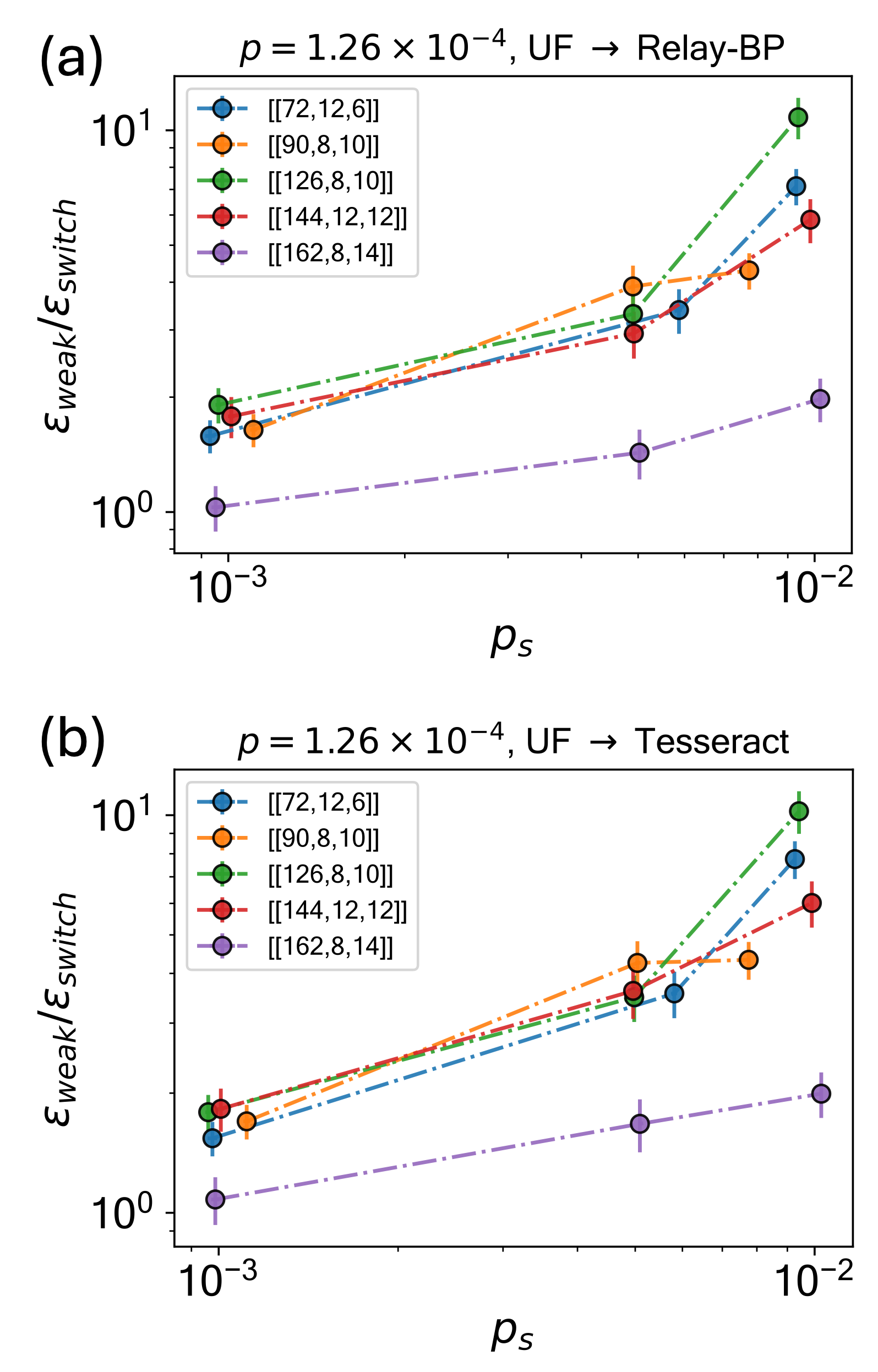}
    \caption{The ratio of the per-cycle logical error rate for switching compared to \texttt{UF} with Pauli-type noise. The physical error rate is $p = 1.26 \times 10^{-4}$. We scan the switching probability with switching to (a) \texttt{Relay-BP} and (b) \texttt{Tesseract}. }
    \label{fig:improvement_uf}
\end{figure}

To evaluate the logical performance gains obtained by the switching protocol compared to the weak decoder, we calculate $ \epsilon_{\text{weak}}/\epsilon_{\text{switch}}$, and consider the physical rate of $p=2\times 10^{-3}$. Figure~\ref{fig:Improvement_bplsd}(a) shows the ratio of LERs per SEC for \texttt{BP-LSD} to \texttt{Relay-BP}  switching, and Fig.~\ref{fig:Improvement_bplsd}(b) for \texttt{BP-LSD} to \texttt{Tesseract} switching. Error bars correspond to the standard deviation of the ratio, where we propagate the standard deviations of the weak and decoder switching protocols.
As the switch rate increases to $10^{-2}$ and beyond,  the performance gains eventually saturate, as we approach the accuracy of the strong decoder. Switching from \texttt{BP-LSD} to \texttt{Relay-BP}  can reduce the LER by a factor up to 20 for $p_s=10^{-2}$, whereas switching from \texttt{BP-LSD} to \texttt{Tesseract} leads to an LER reduction that can exceed three orders of magnitude. The only exception to this improvement is the [[72,12,6]] code, for which the ratio of LERs is close to 2, since the strong decoder does not improve the LER substantially. As the physical error rate decreases further, the LER improvements at a fixed switch rate obtained from decoder switching become even more pronounced than what we observe in Fig.~\ref{fig:Improvement_bplsd} for $p=2\times 10^{-3}$.


In Fig.~\ref{fig:improvement_uf} we consider \texttt{UF} as the weak decoder. We again calculate $\epsilon_{\text{weak}}/\epsilon_{\text{switch}}$ over a range of switch probabilities when we select  \texttt{Relay-BP} [Fig.~\ref{fig:improvement_uf}(a)] or \texttt{Tesseract} [Fig.~\ref{fig:improvement_uf}(b)] as the strong decoder. Both combinations lead to similar LER suppression for the same reasons we mentioned for Fig.~\ref{fig:uf_switching}. While we observe a maximum error suppression of only an order of magnitude, the  performance can be boosted further in the presence of erasures.

While \texttt{UF} and \texttt{BP-LSD} both have a complexity $\mathcal{O}(\kappa^3)$ bounded by the matrix inversion step in cluster verification, where $\kappa$ is the cluster size, both algorithms can obtain close to linear complexity on average in the subthreshold regime~\cite{RoffeNatCommun2025, Sorensenarxiv2024}. In practice, we find that the average cluster size $\kappa$ and algorithm implementation affect runtimes. Further runtime benchmarking is done in Appendix \ref{App:Decoding_times}, with cluster size discussion in Appendix \ref{app:cluster_characterization}.

The gains we reported should be readily achievable for trapped ion architectures given the state-of-the-art millisecond cycle times, combined with the fact that the \texttt{Tesseract} and \texttt{Relay-BP}  decoding time per cycle is within 1 ms for experimentally achievable physical error rates [see Appendix~\ref{App:Decoding_times}]. Our conservative estimates reveal that a switch rate of $p_s=10^{-2}$ is feasible, and can incur LER reductions of even up to 3 orders of magnitude. Different choices of decoders than those we report here could also be used, and might incur even larger improvements. While the choice of \texttt{UF} as a weak decoder for non-topological codes is not so promising under Pauli errors, in several architectures Pauli errors can be converted to erasures with high efficiency. Given lower effective Pauli error rates, \texttt{UF}'s decoding accuracy improves substantially, while retaining a fast decoding speed.

\subsection{Comparison with \texttt{Beamsearch} decoder}

\begin{figure}[!htbp]
    \centering
    \includegraphics[scale=0.76]{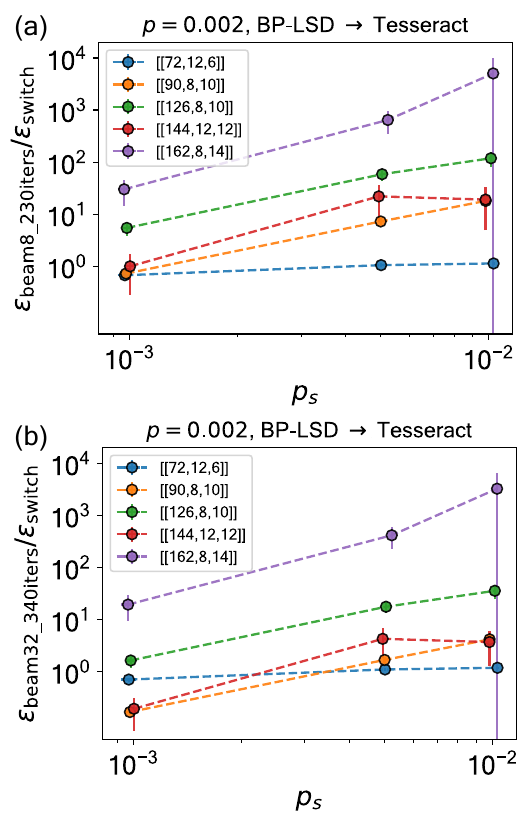}
    \caption{Per-cycle logical error rate ratio of the switching protocol compared to \texttt{Beamsearch} decoder, considering only Pauli-type noise.
    (a) LER ratio of \texttt{Beam8}$\_$\texttt{230iters} to decoder switching protocol as a function of switch rate, when the strong decoder is selected to be \texttt{Tesseract}, and the weak decoder is \texttt{BP-LSD}. The physical error rate is set to $p=2\times 10^{-3}$. (b) Same as in (a), but comparing against \texttt{Beam32}$\_$\texttt{340iters} decoder.}
    \label{fig:Comp_w_Beamsearch}
\end{figure}

Besides looking at the comparison of the decoder switching performance to the weak decoder performance, we also compare against the \texttt{Beamsearch} decoder.  
Specifically, we compare 
our \texttt{BP-LSD} to \texttt{Tesseract} decoder switching protocol with  \texttt{Beam8}$\_$\texttt{230iters} and \texttt{Beam32}$\_$\texttt{340iters}. The parameters for these \texttt{Beamsearch} decoder variants were set according to Ref.~\cite{YePRXQuantum2026}. Further, we consider the Pauli noise model described in Table~\ref{tab:IonicPauli}.

In Fig.~\ref{fig:Comp_w_Beamsearch}(a) we plot the ratio of LER per SEC of \texttt{Beam8}$\_$\texttt{230iters} over the LER per SEC of our protocol for different switch rates. Similarly, Fig.~\ref{fig:Comp_w_Beamsearch}(b) compares the LER performance of our protocol versus \texttt{Beam32}$\_$\texttt{340iters}. We further consider the physical error rate $p=2\times 10^{-3}$. We note that for $p_s=10^{-2}$, which is a feasible switch rate for trapped ions architectures, our protocol can achieve even up to two or three orders of magnitude LER reduction. Given that the CPU decoding latency of \texttt{Tesseract} is similar to \texttt{Beamsearch} for the decoder parameters we considered, and the fact that the \texttt{Tesseract} decoder is not invoked in each shot, the switching protocol provides improved logical performance while reducing the total decoding time [see Appendix~\ref{App:Decoding_times}].


\subsection{Pauli noise accompanied by leakage}
In addition to circuit-level Pauli errors, we investigate the impact of leakage on the decoder switching protocol. Leakage errors are more important for superconducting architectures compared to trapped ions. For superconducting qubits, leakage can be partially counteracted  with low-level control techniques via pulse-shaping (e.g. DRAG) ~\cite{MotzoiPRL2009,GambettaPRA2011,MotzoiPRA2013,HeinsooPRXQ2024}, and higher-level circuit design via leakage reduction units~\cite{MiaoNatPhys2023,GoogleNature2025,PanPRL2025,DiCarloPRXQ2026}, but still remains a challenging noise source.
For trapped ions, leakage events can happen due to imperfect two-qubit gates, or relaxation, but they happen less frequently. 

In what follows, we assume the trapped-ion Pauli noise model with depolarizing rate $p=3\times 10^{-3}$, and a leakage rate $p_{\text{leak}}=10^{-5}$ per syndrome extraction cycle. According to Ref.~\cite{Delfossearxiv2026},
this leakage rate estimates contributions of $1/e$ lifetime decay of metastable states of $^{137}$Ba$^+$ in addition to   unwanted driving of spectator transitions during two-qubit gates. In our circuits, for simplicity, we probabilistically leak every qubit at the beginning of each QEC cycle, which should result in a worst-case performance since  leakage can then propagate through all subsequent two-qubit gates. 

To simulate leakage in addition to Pauli errors, we combine the circuit generation tools from QUITS with the deltakit-stim package~\cite{DeltaKit2025, Chen_2025}, which allows us to sample detection events that arise due to  both Pauli and leakage errors. Deltakit-stim tracks leakage events throughout the circuit and propagates leakage through two-qubit gates. When a qubit is leaked, it is recorded as leaked and sent to the maximally mixed state. If a leaked qubit interacts with another qubit via a two-qubit gate, the latter will also be sent to the maximally mixed state. We construct our circuits with QUITS, and then post-process it by adding leakage instructions ``\texttt{LEAKAGE}$\_$($p_{\text{leak}}$)'' per syndrome extraction cycle, and  ``\texttt{HERALD$\_$LEAKAGE$\_$EVENT}'' instructions which create dedicated measurement records that take the value of 1 if a leakage event happened, or 0 if leakage did not occur. (Note that the measurement records can only be associated with ancilla qubits.)  We associate each of these dedicated measurement records with a leakage detector; the leakage detectors are added on top of the regular detectors in the circuit, and do not compare measurements across rounds. Further, we do not implement leakage-aware decoding, as we want to only benchmark how severe small leakage errors are on our decoder switching protocol. This means that while we sample detection events for all detectors in the circuit, we then discard the leakage-aware detectors, and decode the remaining events using the detector error model constructed from the circuit with $p_{\text{leak}}=0$, and
a leakage-unaware decoder. We leave leakage-aware decoding as an interesting future direction. 

\begin{figure}[!htbp]
    \centering
    \includegraphics[scale=0.7]{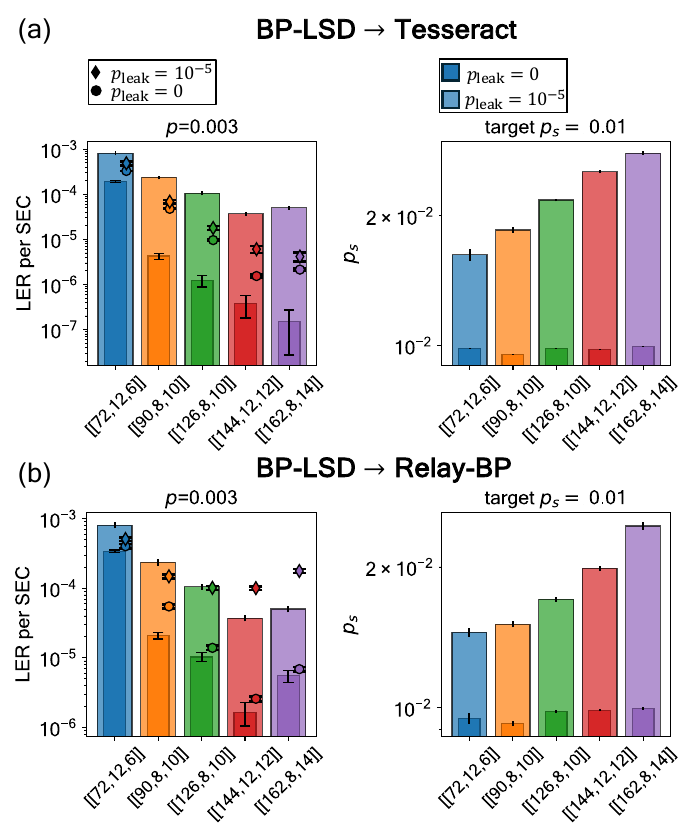}
    \caption{Impact of leakage on the per-cycle logical error rate. (a) Left: logical error rate per syndrome extraction cycle versus BB code, for \texttt{BP-LSD} to \texttt{Tesseract} decoder switching. Dark-colored bars show the performance of \texttt{Tesseract} and light-colored bars of \texttt{BP-LSD} at $p_{\text{leak}}=0$. Circles show the performance of switching at $p_{\text{leak}}=0$, and  diamonds at $p_{\text{leak}}=10^{-5}$. Right: Switch rates for $p_{\text{leak}}=0$ are shown with dark-colored bars, and  for $p_{\text{leak}}=10^{-5}$ with light-colored bars. (b)
    Same as in (a) for \texttt{BP-LSD} to \texttt{Relay-BP}  switching. In both cases, the depolarizing error rate is set to $p=3\times 10^{-3}$.}
    \label{fig:Leakage}
\end{figure}

Due to the presence of leakage the distribution of cluster norms changes. One option would be to re-evaluate the cluster norms in the presence of leakage to find the cutoffs that achieve our target switch rate. Another option is to keep the same cutoffs as in the case of Pauli-only noise, and inspect how the switch rate changes compared to the no-leakage case. We follow this latter approach, using the cluster norm cutoffs that achieve the target switch rate of $p_s\approx 10^{-2}$ for $p_{\text{leak}}=0$. We further select \texttt{BP-LSD} as our weak decoder and exclude \texttt{UF}. The \texttt{UF} decoder is most effective for handling erasures, and thus we do not explore its performance in the presence of leakage.

In Fig.~\ref{fig:Leakage}(a) we show with diamonds the LER per syndrome extraction cycle for $p_{\text{leak}}=10^{-5}$, when we switch from \texttt{BP-LSD} to \texttt{Tesseract}. For comparison, we also show with circles the decoder switching results for $p_{\text{leak}}=0$, as well as the performance of \texttt{BP-LSD} (light-colored bars) and \texttt{Tesseract} (dark-colored bars) at $p_{\text{leak}}=0$. While the presence of leakage deteriorates the maximum performance gains of decoder switching, the switching protocol still preserves substantial gains for all codes. The LER per SEC of the decoder switching protocol still lies within the weak and strong decoder LER bounds at $p_{\text{leak}}=0$. Additionally, as we increase the size of the codes, we see a consistent decrease in the LER, even in the presence of leakage. In Fig.~\ref{fig:Leakage}(a) we further display the actual switch rates we obtain when we configure the cutoffs based on cluster norm distributions collected at $p_{\text{leak}}=0$. Without leakage, the switch rates are close to $10^{-2}$, but in the presence of leakage the switch rates become 2-3 times larger. This is expected, since we fix the same cutoff as in the case of zero leakage. The addition of leakage noise leads to larger clusters, and hence a larger cutoff would be needed to preserve the same rate. As we did not re-evaluate the cluster norms in the presence of leakage, the features we observe are consistent with our expectations. 

The behavior of the decoder switching protocol changes dramatically when we switch from \texttt{BP-LSD} to \texttt{Relay-BP} . These results are presented in Fig.~\ref{fig:Leakage}(b). We find that for the [[72,12,6]] and [[90,8,10]] codes the LER is slightly lower compared to \texttt{BP-LSD} at $p_{\text{leak}}=0$, but for the other codes, the LER is higher compared to \texttt{BP-LSD} at $p_{\text{leak}}=0$. We simulated single sliding window decoding for \texttt{BP-LSD} and \texttt{Relay-BP}  in Appendix~\ref{App:Leakage} and found that \texttt{BP-LSD} is actually more accurate than \texttt{Relay-BP}  when leakage is present. Specifically, the performance of \texttt{Relay-BP}  given $p_{\text{leak}}=10^{-5}$ can deteriorate by up to two orders of magnitude compared to the LER at $p_{\text{leak}}=0$, making it unreliable as a strong decoder and thus eliminating the gains of decoder switching. 

The analysis presented in this section highlights the importance of benchmarking decoders against noise models that go beyond the Pauli approximation. While leakage is treated here via a Pauli approximation, it still reveals the detrimental effects in the logical performance. Testing more refined models  can thus be very important in order to determine whether the decoders can be used for real-time implementation given realistic noise sources that exist in hardware.

\subsection{Pauli noise accompanied by erasures}

\begin{figure}[!htbp]
    \centering
    \includegraphics[width=0.95\linewidth]{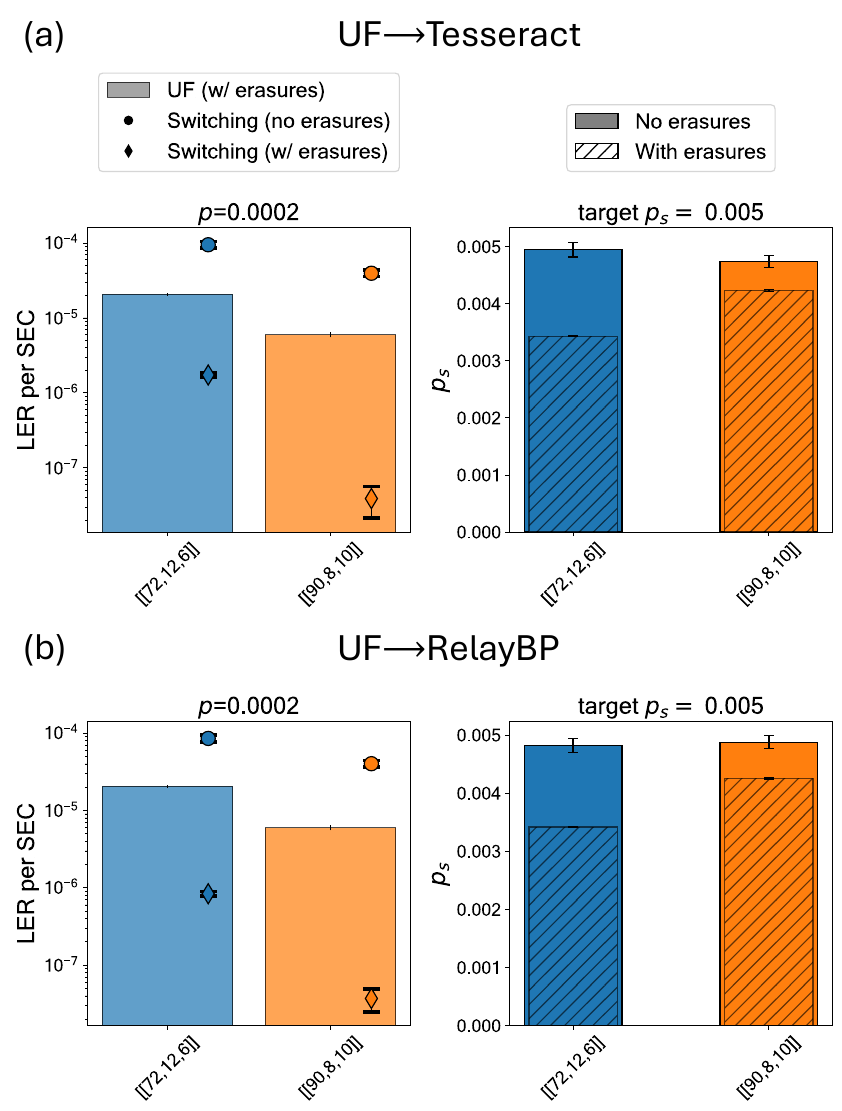}
    \caption{Comparison between switching with erasure conversion rate $R_e = 0.7941$ and erasure-aware \texttt{UF}, and switching without erasure conversion at physical error rate $p = 0.0002$, and switch rate $p_s = 0.005$.  Switching results from (a) are \texttt{UF} $\rightarrow$ \texttt{Tesseract}, and (b) \texttt{UF} $\rightarrow$ \texttt{Relay-BP}. The left plots show switching LERs with and without erasure conversion compared to erasure-aware weak decoding. Decoding with erasure-aware  \texttt{UF} already has lower LER than decoder switching with $R_e=0$. Decoder switching with erasure conversion suppresses the logical error rate additionally by orders of magnitude. The right plots show the switch rates with and without erasures.}
    \label{fig:uf_erasure_benchmarking}
\end{figure}
While qubit leakage is a common problem in many quantum computing architectures, if one is able to determine when it occurs and inform the decoder, decoding performance can be improved. The process of identifying leakage errors and restoring them to the qubit subspace is called erasure conversion. In trapped ions, leakage is detected by monitoring additional transitions on top of the transitions used for gate operations. Once leakage is detected, the population can be pumped back into the qubit subspace, but information about the exact state upon return to the computational space is lost. Thus, the qubit state is assumed to be in a maximally mixed state, signifying the erased information. We note that leakage can be suppressed without harming the data by optical pumping ~\cite{lavasani_quantum_2026, Chen_2025}.

An erasure conversion scheme applicable to trapped ions was introduced in Ref.~\cite{KangPRXQ2023}, which identified experimentally feasible erasure conversion rates for leakage occurring due to two-qubit gates or idling. Since the most dominant errors according to Table~\ref{tab:IonicPauli} are two-qubit errors, and it is possible to convert them into erasures~\cite{KangPRXQ2023},  we convert only those errors into erasures.



The efficiency of converting errors to erasures is quantified by the erasure conversion rate, $R_e$. $R_e$ is the ratio of the erasure rate to the total error rate,  $R_e = \frac{p_\text{erasure}}{p}$. Here, $p=p_\text{erasure}+p_{\text{pauli}}$, where $p$ describes the full error budget for two-qubit errors. While $R_e$ can approach 0.95 for Ca$^+$ meatastable qubits, we consider $R_e = 0.7941$ corresponding to $^{133}\text{Ba}^{+}$ metastable qubits~\cite{KangPRXQ2023}. 

To simulate erasures in addition to Pauli errors, we first determine the probability of erasures based on $R_e$. Given a total two-qubit depolarizing error rate, $p$, the erasure contribution is set to $p_{\text{erase}}=R_e p$, and the Pauli contribution to $p_{\text{pauli}}=(1-R_e)p$. The circuits are created with Stim, using the Pauli noise model of Table~\ref{tab:IonicPauli}, but using instead the modified $p_{\text{pauli}}$ two-qubit depolarizing rate. In addition to this two-qubit error, we sample erasures with probability $p_{\text{erasure}}$ per two-qubit location. This means that we send each qubit involved in the gate to a maximally mixed state. We record a vector of possible error mechanisms, and if an erasure occurred, we set that error mechanism's entry to 1. All other entries are set to 0. This is the vector we  feed to the erasure-aware \texttt{UF} decoder, along with the syndromes. Note that in each shot, the structure of the DEM remains the same, but we need to update the priors, since certain error rates are sent to 0.5 due to the erasure channel.

In Fig.~\ref{fig:uf_erasure_benchmarking} we consider the physical error rate $p=2\times 10^{-4}$, and extract the LER per SEC for the BB codes [[72,12,6]] and [[90,8,10]], under \texttt{UF} decoding, or decoder switching. The performance of the erasure-aware \texttt{UF} decoding is indicated with bars; in this case, the LER reduces by an order of magnitude compared to the case of $R_e=0$. Additionally, circles (diamonds) show the decoder switching performance when $R_e=0$ ($R_e=0.7941$). While we assume a modest switch rate of $p_s=5\times 10^{-3}$, we still obtain appreciable gains of $\epsilon_{\text{weak}}/\epsilon_{\text{switch}}\approx10$ for [[72,12,6]] and $\epsilon_{\text{weak}}/\epsilon_{\text{switch}}\approx 100$ for [[90,8,10]].




\section{Discussion \label{Sec:Discussion}}

We develop a real-time decoding switching protocol for QLDPC codes which toggles between a low-latency ``weak" decoder and a slow ``strong" decoder. We use cluster-based soft information, which allows us to obtain weak decoder confidence for any QLDPC code, and determine decoding windows that need to be handled by the strong decoder. By adjusting the switch rate we tune the decoding accuracy and latency.
Our decoder switching protocol suppresses logical error rates by up to 4 orders of magnitude relative to \texttt{BP-LSD} under the tenable Pauli physical error rate $p = 2 \times 10^{-3}$. Comparing against  the \texttt{Beamsearch} decoder for the same $p$, we find up to 3 orders of magnitude lower logical error rates, and comparable latency. Under a Pauli+leakage error model, we find that \texttt{Tesseract} remains a reliable
strong decoder  and retains decoder switching performance gains, in contrast to \texttt{Relay-BP}. When leakage can be converted to erasures, we find that decoder switching yields up to 3 orders of magnitude of logical error suppression  improvement relative to  \texttt{UF} decoding with zero erasure conversion rate.

While in this work we focused on trapped ion architectures, our protocol could be tailored to any quantum computing architecture. Superconducting qubits, require real-time decoding faster than 1~$\mu$s, which demands FPGA implementations. Easier to implement and lower complexity decoders could be  thus implemented in FPGAs, and supported by strong decoders that run on CPUs. This hybrid decoding could thus boost the logical error suppression, and preserve the real-time decoding latencies imposed by architecture constraints. For trapped ion or neutral atom architectures, the hybrid decoding could be implemented on CPUs or GPUs, with decoders tailored to the accuracy and timing demands of these platforms. The weak decoder does not need to stall while awaiting for the strong decoder's decision as long as appropriate buffer regions before and after the switched windows are formed. Given that the number of windows switched is less than one in a 100, the strong decoder is invoked only a few times in a decoding experiment. Corrections of strong and weak decoders are combined on the boundaries as soon as the decoding processes surrounding the strong decoding windows are finished, as discussed in Ref.~\cite{Fujiiarxiv2025}.

Recently, Ref.~\cite{DelfosseArxiv2026realtime} proposed a dual sliding-window decoder for real-time decoding based on the \texttt{Beamsearch} decoder. This protocol consists of a more accurate ``error decoder'' which runs continuously and attempts to preserve an accurate Pauli frame, and a less accurate ``outcome decoder'' which operates during cat-based logical measurements and is initiated with the Pauli frame of the error decoder. Both decoders operate with fixed small windows and buffer regions, which sacrifice logical accuracy but allow for more immediate processing of windows and a smaller decoding volume. The error decoder in Ref.~\cite{DelfosseArxiv2026realtime} can sometimes exceed the decoding budget, creating a backlog which is subsequently drained by inserting additional syndrome extraction cycles whose role is to allow the decoder to catch up. The outcome decoder is initialized with the Pauli frame of the error decoder, which means if this is not available on time, extra syndrome extraction cycles need to be inserted and the logical measurement is postponed. Our hybrid decoding protocol can straightforwardly substitute the error decoder [\texttt{Beam32}$\_$\texttt{340iters} used in Ref.~\cite{DelfosseArxiv2026realtime}], since it ensures both a higher accuracy (as needed for keeping track of the Pauli frame), but also a better latency compared to \texttt{Beam32}$\_$\texttt{340iters}, given that the strong decoder is infrequently called. Additionally, the strong decoders we considered have better latency compared to \texttt{Beam32}$\_$\texttt{340iters} [see Appendix~\ref{App:Decoding_times}]. Thus, it might be possible that the backlog is much smaller, or non-existent compared to the backlog created by the error decoder. 


As a future work, we believe that hybrid decoding protocols could be developed to address different types of noise. For example, an inner decoder could address leakage~\cite{ZiadNatCommun2025}, and an outer decoder could then decode the remaining errors. Further, different types of soft-output information could be used~\cite{AydingArxiv2026}, that allow the usage of weak decoders that are not cluster-based. Extending this hybrid decoding framework beyond memory experiments, inspecting other choices of decoder combinations and considering further architectural details would be a promising future direction.

\acknowledgments{
The authors would like to thank Sahil Khan, Yingjia Lin, and Bohan Lu for useful discussions and for their feedback on this work. The authors acknowledge support from the National Science Foundation (PHY-
2514847) and the Office of
the Director of National Intelligence (ODNI), Intelligence
Advanced Research Projects Activity (IARPA), under
the Entangled Logical Qubits program through Cooperative Agreement Number W911NF-23-2-0216.

}

\appendix

\section{Decoding time per round for all decoders \label{App:Decoding_times}}

\begin{figure*}[!htbp]
    \centering
    \includegraphics[scale=0.73]{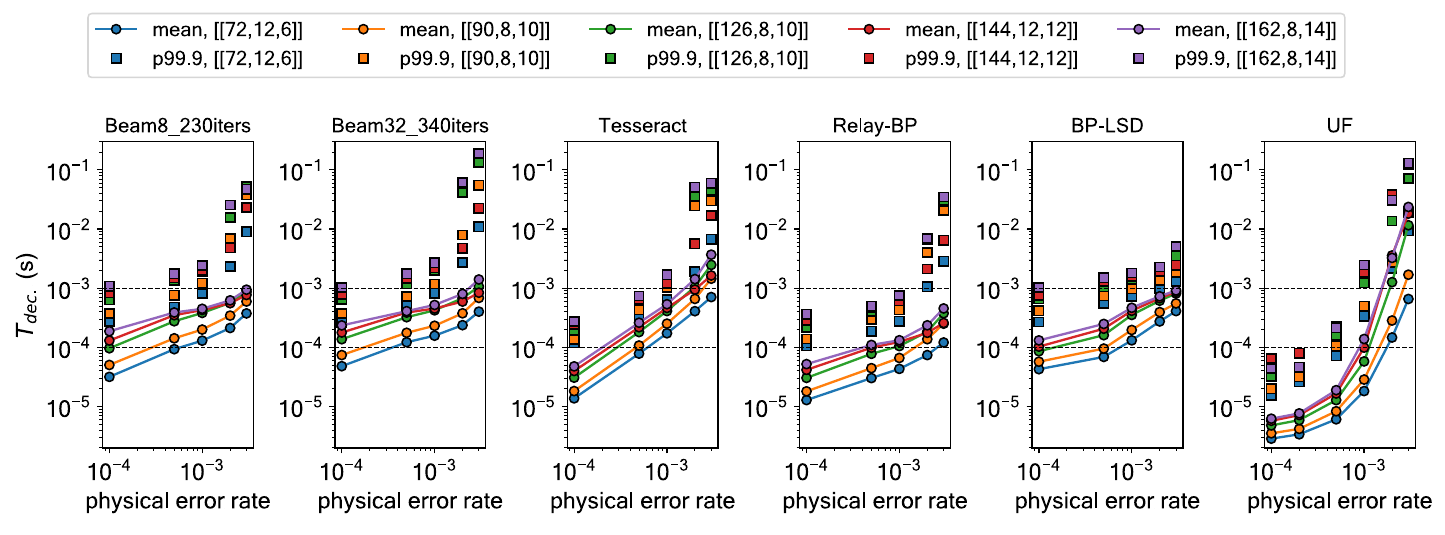}
    \caption{Decoding times per syndrome extraction cycle for the strong decoders considered in this work, versus physical error rate. From left to right we show the performance of \texttt{Beam8}$\_$\texttt{230iters}, \texttt{Beam32}$\_$\texttt{340iters}, \texttt{Tesseract},  \texttt{Relay-BP}, \texttt{BP-LSD} and \texttt{UF}. The circles show average timings, and the squares show the 99.9 percentile. Horizontal dashed lines serve as a guide to the eye, and mark the 0.1 ms and 1 ms decoding times.}
    \label{fig:Timings}
\end{figure*}

In this section we calculate the decoding times per syndrome extraction cycle for the all decoders considered in this work. The timing benchmarks were run on an Apple M5 Pro chip with 24 GB of RAM.  We consider two variants of the \texttt{Beamsearch} decoder namely the \texttt{Beam8}$\_$\texttt{230iters} and \texttt{Beam32}$\_$\texttt{340iters}~\cite{YePRXQuantum2026}, as well as the \texttt{Tesseract},  \texttt{Relay-BP}, \texttt{BP-LSD} and \texttt{UF}  decoders.
Our results are shown in Fig.~\ref{fig:Timings} where we average the timings across 50,000 shots per depolarizing error rate. We use 50,000 shots for all codes and decoders, besides the $p=2\times 10^{-3}$ and $p=3\times 10^{-3}$ for the \texttt{UF}, where we use 200 shots. We set the number of syndrome extraction cycles to $r=25$, and consider full syndrome history decoding. We time only the decode function runtime, and divide the time by the number of cycles, $r$. The circles indicate the average decoding time, and the squares mark the 99.9 percentile runtime. At $p=10^{-4}$, \texttt{Tesseract} and \texttt{Relay-BP}  (which is \texttt{Relay-BP}  1 of Ref.~\cite{Mullerarxiv2025}), are the fastest of all, while the 99.9 percentile runtime of \texttt{Tesseract} is slightly lower than \texttt{Relay-BP} . At $p=10^{-3}$, the average runtime and 99.9 percentile runtime of \texttt{Relay-BP}  are the lowest of all. Note, however, that we have not pushed the decoder parameters of these strong decoders a lot, which would increase latency and accuracy further. At $p=3\times 10^{-3}$, the average runtime of \texttt{Relay-BP}  continues to be the lowest. Overall, given that \texttt{Tesseract} is the most accurate decoder of all, and that for the physical error rates of interest its runtime is comparable with the remaining decoders, makes it a promising choice for a strong decoder. 

Regarding the weak decoders, we find that \texttt{UF} remains very fast for the lowest physical error rates at $10^{-4}$, but its runtime quickly deteriorates for error rates in excess of $10^{-3}$.
While \texttt{BP-LSD} and \texttt{UF} are expected to have similar complexity due to the Gaussian elimination step in the verification of clusters, we see that the \texttt{BP-LSD} runtime is worse than \texttt{UF} at $p\leq 10^{-3}$. We attribute this to non-optimized algorithmic implementation in the package of Ref.~\cite{Roffe_LDPC_Python_tools_2022}. This issue can be resolved by improving the algorithmic implementation. Compared to the strong decoders, the average runtime of \texttt{BP-LSD} is lower than \texttt{Tesseract} for $p\geq  5\times 10^{-4}$, whereas it is comparable or better than \texttt{Beamsearch}. Further, the 99.9 percentile of \texttt{BP-LSD} is lower compared to all strong decoders and lower compared to \texttt{Beamsearch} for the larger error rates.

\begin{figure*}[!htbp]
    \centering
    \includegraphics[scale=1]{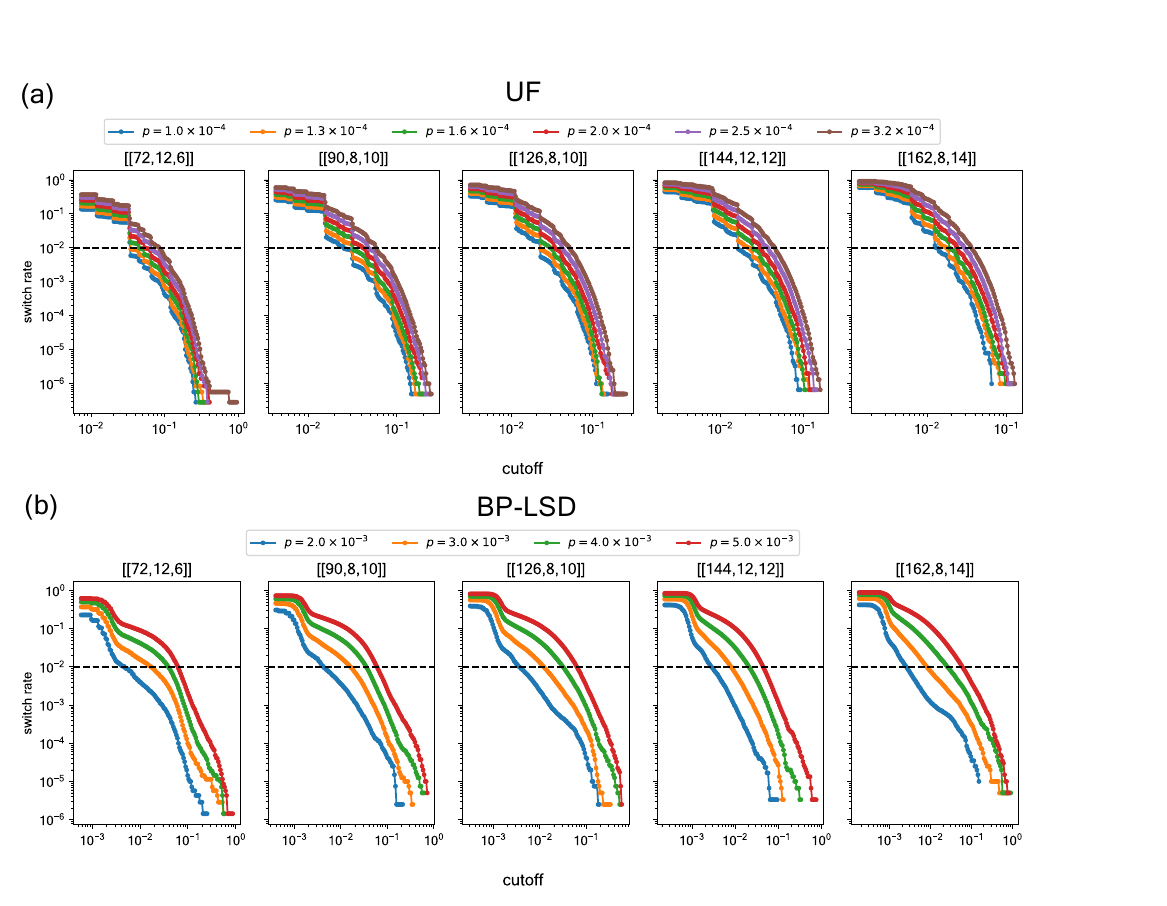}
    \caption{Switch rate distributions versus cluster norm cutoff for varying physical error rates. Distributions for \texttt{UF} and \texttt{BP-LSD} are shown in (a) and (b) respectively. The switch rate of $p_s = 0.01$ is indicated with a horizontal blue line. To determine when to switch, we impose the cluster cutoff for each $p$ according to the desired switch rate. As $p$ increases, so do the cluster norms, so to sustain the desired switch rate, we must increase the cutoff. }
    \label{fig:cluster_cdf}
\end{figure*}
\section{Decoder parameters\label{App:DecoderParams}}
Here we list the parameters that we used for our decoders. For the \texttt{Tesseract} decoder we set:
\begin{itemize}
    \item `det${\_}$beam=10', 
    \item `pq${\_}$limit=50,000', 
    \item `beam${\_}$climbing=True', 
    \item `no${\_}$revisit${\_}$dets=True' .
\end{itemize}
The beam-climbing option means that \texttt{Tesseract} is allowed to search for other det-beams to find a more optimal solution. The no-revisit-dets option allows for faster decoding, possibly at the expense of slightly worse performance. The priority queue (pq-limit) restricts the maximum number of nodes that are in a priority queue during the pathfinding search.

For \texttt{Relay-BP} we set the parameters:
\begin{itemize}
    \item `gamma${\_}$dist${\_}$interval=(-0.175,0575)',
    \item `gamma0=0.125',
    \item `pre${\_}$iter=80',
    \item `num${\_}$sets=300',
    \item `set${\_}$max${\_}$iter=60',
    \item `stop${\_}$nconv=1'.
\end{itemize}
The gamma parameters are optimal for the Gross code $[[144,12,12]]$ according to Ref.~\cite{Mullerarxiv2025}. The number of sets corresponds to the total number of legs (realizations) for \texttt{Relay-BP} 1 based on Ref.~\cite{Mullerarxiv2025}, with a maximum number of 60 iterations per leg. We further abort the search for more optimal solutions, since convergence stops if a single solution is found.

Finally, for \texttt{BP-LSD} we set:
\begin{itemize}
    \item `bp${\_}$method=product${\_}$sum',
    \item `max${\_}$bp${\_}$iters=30',
    \item `schedule=parallel',
    \item `lsd${\_}$method=lsd${\_}$cs'
    \item `lsd${\_}$order=0'.
\end{itemize}

\section{Cluster Distribution Characterization \label{app:cluster_characterization}}

We characterize the cluster distributions for each code by computing a histogram of the cluster norms for each physical probability $p$. We take $r=25$ rounds to collect the cluster norm distributions with sliding window decoding, where we set $F=d/2$ and $n_{\text{buffer}}=d$. The total number of shots is set to 100,000 for \texttt{BP-LSD} and 500,000 for \texttt{UF}. By imposing a cluster norm cutoff $q_{\text{th}}$, we calculate the probability that switching events occur, based on the cumulative distribution function of cluster norms. For a desired switch rate, it is possible to directly find the necessary cluster norm cutoff for each code and physical error rate $p$. The results are shown in Fig. \ref{fig:cluster_cdf}(a) for \texttt{UF} and in Fig. \ref{fig:cluster_cdf}(b) for \texttt{BP-LSD}.

In Fig. \ref{fig:cluster_cdf}, we immediately notice that the \texttt{UF} clusters produce less distinct switch rate distributions at different physical $p$. This is because the \texttt{UF} clusters grow more quickly and uniformly for each $p$. This is reflected on the fact that given the same switch rate for both \texttt{UF} and \texttt{BP-LSD}, the \texttt{UF} cluster norm cutoffs are larger.

The differences between \texttt{UF} and \texttt{BP-LSD} are due to their algorithmic cluster growth strategy. Unweighted \texttt{UF} grows clusters as a breadth first search, while \texttt{BP-LSD} intentionally adds edges according to their weight. Clusters sizes produced by \texttt{BP-LSD} can be calculated analytically by formulating a percolation problem \cite{RoffeNatCommun2025}. Unweighted \texttt{UF} clusters grow uniformly dependent on the node degree. While we expected \texttt{UF} runtimes to reflect the matrix inversion required on larger cluster sizes, this effect is only visible at higher $p$, as we found in Appendix~\ref{App:Decoding_times}.

\section{Additional leakage simulations \label{App:Leakage}}
\begin{figure}[!hbtp]
    \centering
    \includegraphics[scale=0.65]{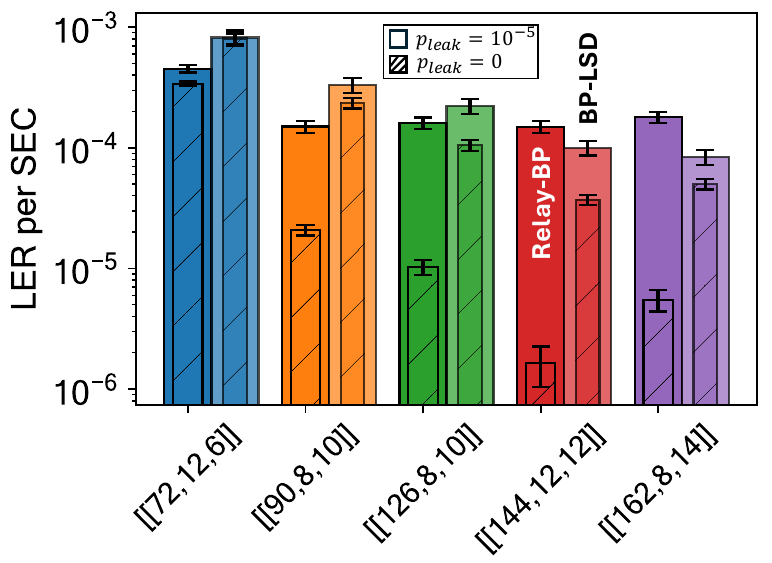}
    \caption{Logical error rate per syndrome extraction cycle for \texttt{Relay-BP}  and \texttt{BP-LSD} sliding window decoding, for five BB codes. The darker bars show the performance of \texttt{Relay-BP} , and lighter bars the performance of \texttt{BP-LSD}. The hatched bars correspond to zero leakage, and the solid bars to $p_{\text{leak}}=10^{-5}$.}
    \label{fig:Relay_and_bplsd_w_leakage}
\end{figure}

Here we show the performance of \texttt{Relay-BP}  and \texttt{BP-LSD} for a depolarizing rate of $p=3\times 10^{-3}$ and $p_{\text{leak}}=10^{-5}$. The leakage error is applied as in the main text; we consider leakage channel per qubit at the start of each QEC cycle. We consider sliding window decoding for $r=25$ syndrome extraction cyles, and set $F=d/2$ and $n_{\text{buffer}}=d$. The logical error rate per syndrome extraction cycle is shown in Fig.~\ref{fig:Relay_and_bplsd_w_leakage}. The darker bars correspond to the performance of \texttt{Relay-BP}  and the lighter bars to the performance of \texttt{BP-LSD}. Further, the hatched bars correspond to $p_{\text{leak}}=0$, and the solid bars to $p_{\text{leak}}=10^{-5}$. We note that the performance of \texttt{BP-LSD} deteriorates only by a few times in the presence of leakage. For \texttt{Relay-BP},  however, the presence of leakage can deteriorate its performance by even two orders of magnitude (e.g., for [[144,12,12]]), making it even less accurate than \texttt{BP-LSD}.

\clearpage
\nocite{*}

\bibliography{apssamp}

\end{document}